\documentclass[iop]{emulateapj}
\usepackage{graphicx}

\def\simgr{\,\hbox{\hbox{$ > $}\kern -0.8em \lower 1.0ex\hbox{$\sim$}}\,}
\def\simle{\,\hbox{\hbox{$ < $}\kern -0.8em \lower 1.0ex\hbox{$\sim$}}\,}

\newcommand{\fermi}{{\it Fermi}}
\newcommand{\chandra}{{\it Chandra}}
\newcommand{\xmm}{{\it XMM-Newton}}
\newcommand{\mspone}{PSR J1628$-$3205}
\newcommand{\msptwo}{PSR J1048$+$2339}
\newcommand{\mspthree}{XMMU J083850.38$-$282756.8}
\newcommand{\fgl}{3FGL J0838.8$-$2829}

\shortauthors{Cho et al.}
\shorttitle{Variable Heating in Redbacks}

\begin{document}
\title{Variable Heating and Flaring of Three Redback Millisecond Pulsar Companions}

\author{Patricia B. Cho, Jules P. Halpern, and Slavko Bogdanov}
\affiliation{Department of Astronomy, Columbia University, 550 West 120th Street, New York, NY 10027-6601, USA; patricia.cho@columbia.edu}

\begin{abstract}
  We are monitoring established and putative redback millisecond pulsars
  (MSPs) in time-series photometry, repeatedly covering their 5--6~hr
  orbital light curves in $r^{\prime}$ or $R$.  On timescales of months,
  \msptwo\ and \mspthree\ exhibit similar variability of
  $\approx 0.3$ mag on the heated side of the companion star.
  However, the heating light curve is rarely symmetric, suggesting that
  the intrabinary shock generated by the pulsar wind is skewed in addition to
  being variable, or that changing magnetic fields intrinsic to the companion
  channel the pulsar wind.  In addition to this variable heating, there are
  long-lived flaring states that increase the brightness by an additional
  0.5~mag, with variability on $\approx 10$ minute timescales.  These flares
  also appear to originate on the heated side of the companion,
  while the ``night''-side 
  brightness remains relatively stable. Somewhat less active, \mspone\ has an
  optical light curve that is dominated by tidal distortion
  (ellipsoidal modulation), although it too shows evidence of variable and
  asymmetric heating due to shifting magnetic fields or migrating star spots.
  These effects frustrate any effort to derive system parameters such as
  inclination angle and Roche-lobe filling factor from optical light curves
  of redback MSPs.  We also report on two \chandra\ X-ray observations of
  \msptwo\ that show strong orbital modulation, possibly due to beaming
  along the intrabinary shock, and a third observation that is dominated
  by flaring.  The peak flare luminosity in the 0.3--8 keV band
  is $\approx12\%$ of the pulsar's spin-down power, which may require
  magnetic reconnection.  None of these three systems has yet shown a
  transition back to an accreting state.
\end{abstract}


\section{Introduction} \label{sec:intro}

The Large Area Telescope on the {\it Fermi Gamma-ray Space Telescope}
is making a major contribution to the science of millisecond pulsars (MSPs).
Of the 216 $\gamma$-ray pulsars detected to date, 99 are MSPs\footnote{https://confluence.slac.stanford.edu/display/GLAMCOG/\\Public+List+of+LAT-Detected+Gamma-Ray+Pulsars}.
The unique conditions under which MSPs are formed from their
low-mass X-ray binary (LMXB) progenitors, and the continued episodic accretion
that some of them display, provide new opportunities to explore interesting
problems such as the nature of propeller accretion and the maximum mass of
a neutron star.

MSPs are neutron stars with spin periods even faster than those of young pulsars.
The recycling scenario is now widely understood to be the mechanism for their ``spin-up''.
An accretion disk fed by material stripped from a Roche lobe filling companion
transfers angular momentum to the neutron star.  In this way the neutron star spin
periods are ``recycled'' into the millisecond regime \citep{alp82}.  Black widows
(BWs) and redbacks comprise the subclass of MSPs whose binary orbits are $\simle$1~day and
whose radio pulses are often periodically
eclipsed. An intrabinary shock driven by the relativistic pulsar wind produces a
secondary wind of ablated plasma trailing off of the companion star \citep{phi88}.
While the cross-sectional area of the companion star alone would be insufficient to
eclipse the signals for the durations observed, the companion's plasma wind is
extensive enough to disperse and absorb
the radio pulsations as the neutron star reaches superior conjunction.

BW companion masses range from $\approx0.01-0.05$ $M_{\odot}$ and redback companions range
from $\approx0.1-0.5$ $M_{\odot}$. The compact binary orbits often tidally distort
their companion stars, which is evident as ellipsoidal orbital modulation.
The optical light curves of many BW and redback systems also
show heating on the side of the companion reaching in toward the inner
Lagrange point \citep{rom11,kon12,bre13,gen14,bog11,bog14a,bog14b}. This phenomenon,
as well as orbitally modulated, power-law X-ray emission,
lends further credence to the presence of an intrabinary shock. The amplitude of the
modulation is proportional to the inclination of the binary system's orbital plane,
with the maximum possible modulation
corresponding to an inclination angle of $90^{\circ}$.

There are over 60 BW and redback pulsars known,\footnote{https://apatruno.wordpress.com/about/millisecond-pulsar-catalogue/}
almost equally divided between globular clusters and the Galactic field.
Most of them were discovered in follow-ups of
\fermi\ sources, which preferentially detect these subclasses of MSPs.
In addition, there are several putative BWs or redbacks that are very
likely counterparts of \fermi\ sources (see \citealt{li18} for a list). 
The systems we investigate here are all \fermi\ sources: two previously
confirmed redbacks exhibiting some evidence of heating,
\mspone\ and \msptwo\ \citep{li14,den16}, and one
putative redback, \mspthree\ identified with \fgl\ \citep{hal17b}.
The goals of these observations are to characterize long-term variations in
pulsar wind heating of the companions, and to search for flaring.
In addition to these three variable objects, we have several observations
of three more redbacks that show little if any change; these will be reported
separately. Section~2 describes our observations and data reduction
methodology.  In Sections 3--5 we present new light
curves and compare them with previous observations. Section~6 discusses the
implications of the changes in the light curves, and we present our conclusions
in Section~7.

\section{Observations}
\begin{deluxetable}{cclcr} 
\tablecolumns{5} 
\tablewidth{0pt} 
\tablecaption{Log of Time-Series Photometry} 
\tablehead{\colhead{Telescope} & \colhead{Filter} & \colhead{Date (UT)}
& \colhead{Time (UTC)} & \colhead{Phase ($\phi$)\tablenotemark{a}}}
\startdata
\cutinhead{\msptwo}
1.3 m & $R$ & 2016 Jan 10  & 05:51--10:01 & 0.31--0.98 \\
1.3 m & $R$ & 2016 Jan 12  & 05:59--13:30 & 0.32--1.56 \\
1.3 m & $V$ & 2016 Jan 13  & 05:49--13:33 & 0.28--1.56 \\
1.3 m & $R$ & 2016 Mar 4   & 03:03--08:40 & 0.41--1.31 \\
1.3 m & $R$ & 2016 Mar 5   & 04:16--08:59 & 0.60--1.37 \\
1.3 m & $R$ & 2016 Mar 7   & 02:29--08:40 & 0.28--1.30 \\
1.3 m & $R$ & 2016 May 29  & 03:41--06:19 & 0.78--1.20 \\
1.3 m & $R$ & 2016 May 31  & 03:23--06:12 & 0.71--1.16 \\
1.3 m & $R$ & 2016 Jun 3   & 03:20--05:58 & 0.67--1.10 \\
2.4 m & $r^{\prime}$ & 2016 Jun 7 & 03:29--05:48 & 0.67--1.04 \\
1.3 m & $R$ & 2016 Dec 24  & 07:35--13:18 & 0.70--1.64 \\
1.3 m & $R$ & 2017 Feb 27  & 02:35--07:03 & 0.34--1.07 \\
1.3 m & $R$ & 2017 Mar 24 & 03:06--10:05 & 0.22-1.34 \\
1.3 m & $V$ & 2017 Mar 25 & 02:43--09:09 & 0.15--1.20 \\
1.3 m & $r^{\prime}$ & 2018 Jan 12 & 06:43--13:25 & 0.38--1.48 \\
1.3 m & $r^{\prime}$ & 2018 Jan 14 & 06:57--13:29 & 0.41--1.48 \\
1.3 m & $r^{\prime}$ & 2018 Feb 18 & 05:42--13:06 & --0.86--1.10 \\
1.3 m & $r^{\prime}$ & 2018 Feb 22 & 07:43--13:05 & 0.21-1.10 \\
1.3 m & $r^{\prime}$ & 2018 Apr 19 & 02:43--09:09 & --0.09--0.97 \\
1.3 m & $r^{\prime}$ & 2018 Apr 20 & 02:41--08:51 & --0.10--0.99 \\
1.3 m & $R$ & 2018 Apr 21 & 02:34--09:00 & --0.13--0.93 \\
1.3 m & $r^{\prime}$ & 2018 May 20 & 03:34--07:00 & 0.78--1.34 \\
1.3 m & $r^{\prime}$ & 2018 May 21 & 03:15--06:03 & 0.72--1.18 \\
\cutinhead{\mspthree}
2.4 m & $r^{\prime}$ & 2016 Dec 27 & 07:02--12:41 & 0.04--1.12 \\
2.4 m & $r^{\prime}$ & 2017 Feb 25 & 02:57--08:20 & $-0.05$--0.98 \\
1.3 m & $r^{\prime}$ & 2017 Nov 17 & 09:36--12:57 & 0.52--1.16 \\
1.3 m & $r^{\prime}$ & 2017 Nov 19 & 09:41--13:00 & $-0.13$--0.50 \\
1.3 m & $r^{\prime}$ & 2017 Dec 14 & 08:01--13:17 & 0.09--1.10 \\
1.3 m & $r^{\prime}$ & 2017 Dec 16 & 07:38--12:54 & 0.33--1.34 \\
1.3 m & $r^{\prime}$ & 2018 Jan 13 & 08:52--11:18 & 0.10--0.56 \\
1.3 m & $r^{\prime}$ & 2018 Feb 23 & 03:06--08:21 & 0.11--1.11 \\
1.3 m & $r^{\prime}$ & 2018 Mar 16 & 02:23--07:16 & $-0.14$--0.79\\
\cutinhead{\mspone}
2.4 m & $R$ & 2014 May 26 & 05:58--09:36 & 0.88--1.60 \\
2.4 m & $R$ & 2014 May 27 & 06:04--09:37 & 0.70--1.40 \\
2.4 m & $R$ & 2014 May 28 & 05:34--09:28 & 0.41--1.18 \\
2.4 m & $R$ & 2017 May 29 & 05:12--09:59 & 0.72--1.66 \\
2.4 m & $R$ & 2018 May 20 & 05:37--10:38 & 0.15--1.14
\enddata
\tablenotetext{a}{Orbital phase zero corresponds to the ascending node of the pulsar.}
\label{tab:optlog}
\end{deluxetable}

We obtained new time-series photometry on 33 nights between 2016 Jan and 2018 May.
All observations were performed using MDM Observatory's 1.3~m McGraw-Hill Telescope
or 2.4~m Hiltner Telescope on Kitt Peak. For observations on the 1.3~m, we used the
thinned, backside-illuminated SITe CCD ``Templeton'' which has a plate scale of
$0^{\prime\prime}\!.509$ pixel$^{-1}$. For observations on the 2.4~m, we used the Ohio State
Multi-Object Spectrograph \citep[OSMOS,][]{mar11} in imaging mode. The OSMOS plate
scale is $0^{\prime\prime}\!.273$ pixel$^{-1}$. Most images were taken in the $R$
or $r^{\prime}$ filters.
CCD readouts for observations done using the 2.4~m were binned to
reduce dead-time, as this could be done without compromising the quality of the
photometry. Exposure times were all 300~s except for data taken in the $V$ filter
for which exposure times were 360~s. Dead-time was 26~s for all exposures
except for the binned 2.4~m images where dead-time was 13~s. A log of the
observations is given in Table~\ref{tab:optlog}.

We used standard IRAF routines to reduce the images. We used \textit{zerocombine}
and \textit{flatcombine} each in conjunction with \textit{ccdproc} in the noao.imred.ccdred
package to process the bias and twilight flat-field images. We used \textit{ccdproc} to
subtract the bias from and flatten the science images.
We performed differential time-series photometry using \textit{phot} in the noao.digiphot.daophot package. The
parameters in \textit{phot} were optimized based on the full width half maximum (FWHM) of the PSF for the pulsar companion and comparison stars. 
We also performed differential
photometry on the comparison stars for each set of exposures against secondary comparisons
to ensure they are not themselves variable. The same set of comparison and secondary
comparison stars were also measured across different sets of exposures to confirm
the absence of any long term variability.
Absolute photometric
calibration of comparison stars for \mspone\ was performed using a set of observations of
Landolt (1992) standard stars taken on 2017 June 2. Magnitudes of comparison stars for \msptwo\ and \mspthree\
were taken from SDSS and, where needed, converted to $V$ and $R$ using Lupton (2005) transformations\footnote{http://classic.sdss.org/dr4/algorithms/sdssUBVRITransform.html}.

Previously published data from 2014 on \mspone\ \citep{li14}
were reextracted; differential photometry was conducted using
the same set of comparison stars used previously
and calibrated to the same secondary standards as the new data.
There are some minor discrepancies between the light curves presented here
and those from \citep{li14}, presumably due to slight differences in the
parameters used to extract the photometry.

Figures in the following sections plot calibrated magnitude as a function
of orbital phase for each set of exposures, where $\phi=0$ corresponds
to the ascending node of the pulsar.  We first corrected each exposure to
Barycentric Dynamical Time (TDB).  To determine the orbital phase for the
two radio pulsars, we calculated the number of orbital periods that had
elapsed from the time of ascending node to the midpoint of the exposure
using the radio pulsar ephemerides.  For \mspthree\ we used the data themselves
to fit a revised orbital ephemeris using the method of \cite{hal17b}, in
which a fit to the minimum of the light curve was used to define the
epoch assumed to be the inferior conjunction of the companion ($\phi=0.25$).
A few outlying points were deleted because they were affected by cosmic-ray
contamination, and a few others that had very large errors were not used.

\begin{figure*}
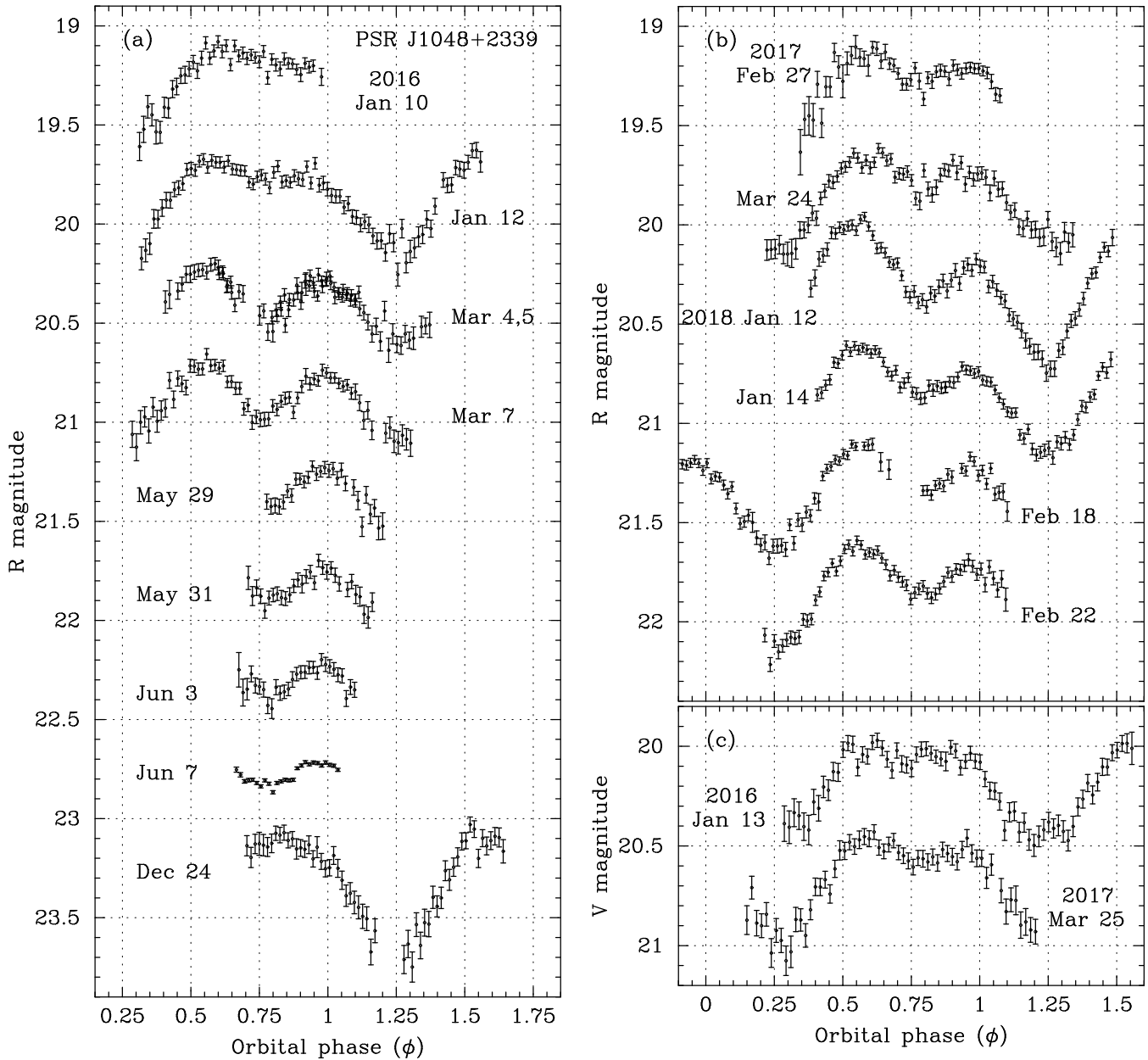

\begin{minipage}[b]{0.5\textwidth}
\includegraphics[width=8.7cm]{f1a.eps}
\end{minipage}
\begin{minipage}[b]{0.5\textwidth}
\includegraphics[width=8.7cm]{f1b.eps}
\vspace{0.17cm}
\includegraphics[width=8.7cm]{f1c.eps}
\end{minipage}
\vspace{-0.1in}
\caption{Optical light curves of \msptwo\ as a function of orbital phase.
 A log of the observations is given in Table~\ref{tab:optlog}.
 In each panel data sets below the uppermost one
 are displaced downward by a multiple of 0.5 mag for clarity.
}
\vspace{0.2in}
\label{fig:image2}
\end{figure*}

\begin{figure}
\centerline{
\includegraphics[angle=0,width=1.\linewidth,clip=]{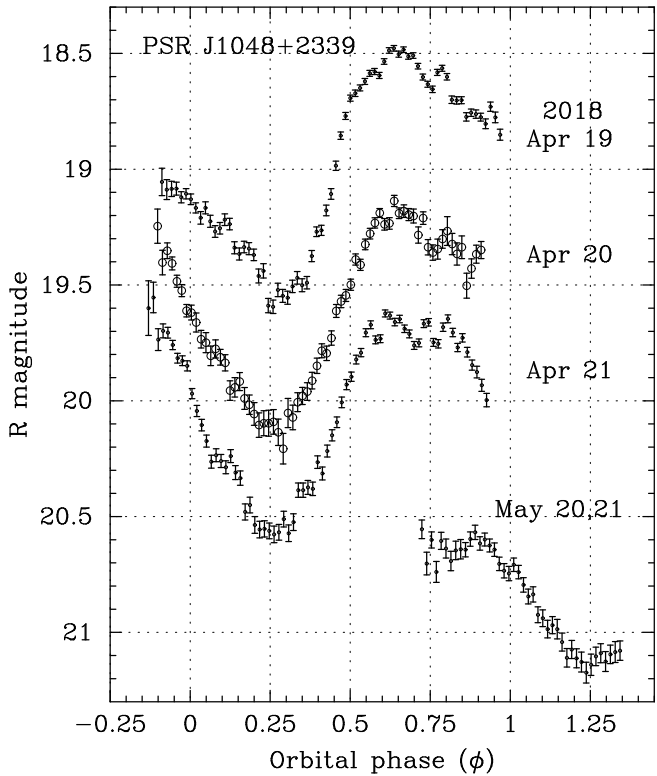}
}
\caption{Optical light curves of \msptwo\ in its flaring state in
 2018 April, and having returned to the quiescent level in May.
 A log of the observations is given in Table~\ref{tab:optlog}.
 Each data set after April 19 is
 displaced downward by a multiple of 0.5 mag for clarity.
 The different symbols aid in the separation of
 closely spaced light curves.
}
\label{fig:flare}
\end{figure}

\section{\msptwo}

\subsection{Optical Light Curves}

Figures~\ref{fig:image2} and \ref{fig:flare} show light curves of
\msptwo\ folded on the radio pulsar ephemeris of \citet{den16}.
Specifically, we use the (6.01~hr) orbital period and time of ascending node,
but not the period derivative(s), which are not predictive
on long timescales.  Our earliest data from 2016 January
(Figure~\ref{fig:image2}a) resemble the mean light curve in the
discovery paper \citep{den16}, which was extracted from the 2005--2013 
Catalina Real-Time Transient Survey (CRTS).  In particular,
there is a minimum as expected at $\phi=0.25$, the inferior conjunction
of the secondary, and a downward sloping plateau from $0.5<\phi<1.0$ 
that is evidently due to pulsar heating of the companion, but not exactly
symmetric about the expected $\phi=0.75$.  Such effects were previously
noted and interpreted as asymmetric heating in other redback systems
\citep{wou04,li14,hal17b}. 

There subsequently appeared a significant change at $\phi=0.75$,
the expected phase of maximal heating.  By 2016 March the brightness
at this phase dipped by as much as 0.4 magnitudes below the maximum at 
$\phi=0.5$, leaving two clearly defined peaks that are the signature
of ellipsoidal modulation.  In the absence of any heating, ellipsoidal
modulation would be characterized by two equal maxima at $\phi=0$
and $\phi=0.5$, while the minimum at $\phi=0.75$ would be deeper than
that at $\phi=0.25$ due to gravity darkening. While heating was still
evident because the brightness at $\phi=0.75$ remained 
higher than at $\phi=0.25$, heating was no longer dominant over tidal
distortion in shaping the light curve.  Seasonal visibility restrictions
then prevent us from getting complete orbital coverage, but by
2016 December 24 it appeared that heating had returned to its initial
strength at $\phi=0.75$.

Figure~\ref{fig:image2}b shows additional light curves obtained in
2017 and 2018, which display a variety of levels of heating.  These
states are characterized by varying depths of the local minimum
at $\phi=0.75$, and also by significant changes in the relative
heights of the two peaks at $\phi=0.5$ and $\phi=1.0$.  While
major changes appear on time scales of months, small variations
can also take place in days, such as in the peak brightness
between 2018 January 12 and 14.

Close examination reveals that the minimum expected at $\phi=0.75$ doesn't
always fall exactly at this phase, but after 2016 March
is delayed by $\approx0.05$.   This shift is unlikely to be due to a
drift in the orbital ephemeris because the phase of the primary 
minimum (viewing the dark side of the companion) remains consistent
with $\phi=0.25$ throughout.  It may be the result of a number of 
different physical mechanisms.  First, the skewed intrabinary
shock that causes the sloping maxima and asymmetric peaks may
also move the phase of minimum. 
Second, the companion may have a large starspot that occupies a
position near the inner Lagrange point.  Third, a shifting magnetic
field intrinsic to the companion may channel the pulsar wind to
a different location on its surface. 

\begin{figure*}
\includegraphics[angle=0,width=\textwidth]{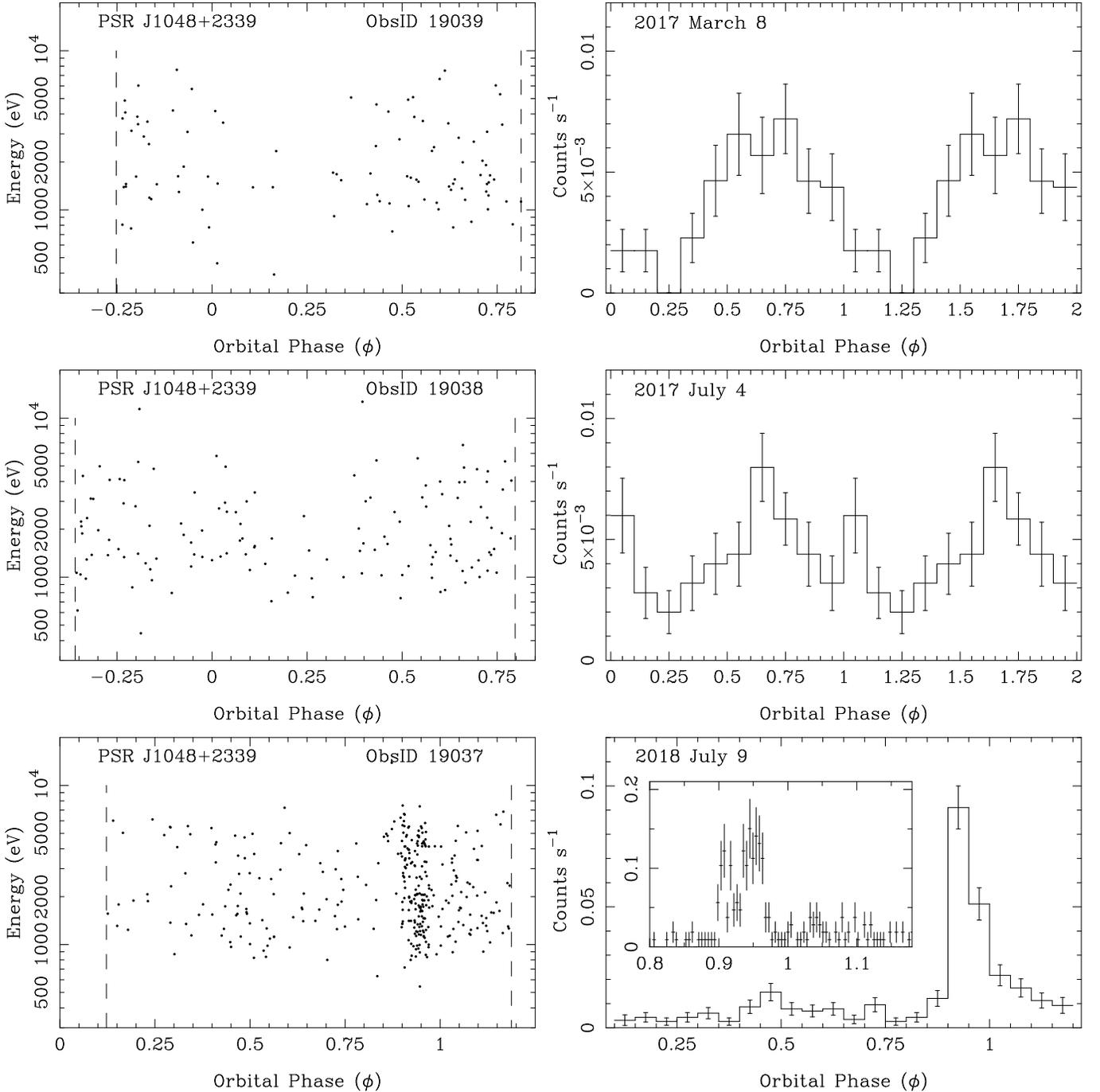}
\caption{\chandra\ light curves of \msptwo\ from three observations.
The left-hand panels show the orbital phase and energy of each photon,
with the vertical dashed lines marking the start and end of the observation.
Each observation spans slightly more than one pulsar orbit.
The right-hand panels are the binned light curves corrected for exposure
time at each orbital phase.  In the first two observations the
binned light curve is repeated for clarity.  The third observation
is dominated by flares.  The inset shows the main flare in 100~s bins.
}
\label{fig:xray}
\end{figure*}

We obtained only two nights of data in the $V$ filter for \msptwo,
in 2016 and 2017.  These two light curves closely resemble the $R$-band
data at the same epochs, and are reminiscent of the transitional MSP 
J1023+0038 in its radio pulsar state \citep{wou04,tho05}.

In 2018 April we observed a new, flaring state
for three consecutive nights (Figure~\ref{fig:flare}).
One month later, the light curve had returned to the
quiescent level.  The flaring
state is characterized by up to an additional
0.5 mag brightening on the heated side of the star,
which previously had $R\simgr19.0$,
but now reached up to $R=18.5$.  In addition, rapid
variability on timescales of 10 minutes is evident.
It is possible that faster variations could be detected
with higher time resolution data. In addition, there
may be rapid variability during some of quiescent periods
in Figure~\ref{fig:image2}, but this is less certain.

The existence of occasional
flaring states was inferred in \citet{den16} from the
couple of dozen observations out of the $\approx 400$
from the CRTS that were up to a magnitude brighter than
the average plateau. It appears that the location of the
flares, as well as the more general long-term variability,
are confined to side of the companion facing the pulsar,
because the magnitude at $\phi=0.25$, centered on the
dark side, has remained relatively stable throughout
our monitoring.  We see that always $R\approx19.6$
at $\phi=0.25$, with a maximum range of $\pm0.1$.
Any systematic variation due to the calibration process
is limited to $\pm0.02$ mag, so $\pm0.1$ mag is a conservative
upper limit on the range of intrinsic stellar variability.
Thus, the much larger change of up to 1 mag at other phases
is not intrinsic to the companion star alone,
but is related to and requires the influence of the pulsar.

\begin{deluxetable*}{lcccccccc}
\tablecolumns{9}
\tablewidth{0pt}
\tablecaption{\chandra\ X-ray Spectral Fits for PSR J1048+2339\label{tab:spec}}
\tablehead{\colhead{ObsID} & \colhead{Date} & \colhead{Exposure} & \colhead{Counts} &
\colhead{Rate\tablenotemark{a}} & \colhead{$N_{\rm H}$} & \colhead{$\Gamma$} & \colhead{$F_X$\tablenotemark{b}} &
\colhead{$\chi^2$/dof} \\
\colhead{} & \colhead{(UT)} & \colhead{(s)} &  & \colhead{(s$^{-1}$)} & \colhead{($10^{20}$ cm$^{-2}$)} & \colhead{} &
\colhead{(0.3--8 keV)} & \colhead{}}
\startdata
19039 & 2017 March 8 & 22,551 & 98 & $0.0039(4)$ & $4^{+69}_{-4}$ & $1.56^{+1.34}_{-0.56}$ & $5.1^{+6.5}_{-1.3}$ &  $3.02/3$ \\
  & & & & & $2.5$\tablenotemark{c} & $1.56^{+0.49}_{-0.50}$ & $5.1^{+1.5}_{-1.3}$ & $3.02/4$ \\
19038 &  2017 July 5 & 24,729 & 136 & $0.0044(4)$ & $6^{+43}_{-6}$ & $1.54^{+0.78}_{-0.41}$ & $7.4^{+3.1}_{-1.6}$ & $2.11/6$ \\
  & & & & & $2.5$\tablenotemark{c} & $1.46^{+0.36}_{-0.36}$ & $7.3^{+1.6}_{-1.5}$ & $2.14/7$ \\
\hline
19038+19039 & 2017 March, July & 47,280 & 234 & $0.0042(3)$ & $<20$ & $1.47^{+0.43}_{-0.29}$ & $5.9^{+1.1}_{-1.0}$ & $9.42/12$ \\
            & & & & & $2.5$\tablenotemark{c} & $1.51^{+0.31}_{-0.30}$ & $5.9^{+1.0}_{-1.0}$ & $9.63/13$ \\
\hline
19037 &  2018 July 9 & 22,753 & 344 & $0.0151(8)$ & $<17.5$ & $1.23^{+0.27}_{-0.19}$ & $23.0^{+2.6}_{-2.5}$ &  $19.27/19$ \\
      &              &        &     &             &  $2.5$\tablenotemark{c} & $1.26^{+0.19}_{-0.19}$ & $23.0^{+2.6}_{-2.5}$ & $19.49/20$ \\
19037 flare\tablenotemark{d} &  2018 July 9 & 7,162 & 243  & 0.0339(22) & $<24.7$ & $1.18^{+0.38}_{-0.23}$ & $51.9^{+7.5}_{-6.9}$ & $12.51/12$ \\
      &              &        &     &             &  $2.5$\tablenotemark{c} & $1.22^{+0.23}_{-0.24}$ & $52.0^{+7.4}_{-6.8}$ & $12.62/13$
\enddata
\tablenotetext{a}{Phase-averaged count rate and uncertainty.}
\tablenotetext{b}{Unabsorbed 0.3-8 keV flux in units of $10^{-14}$ ergs cm$^{-2}$ s$^{-1}$.}
\tablenotetext{c}{$N_{\rm H}$ held fixed at the Galactic value from \citet{kal05}.}
\tablenotetext{d}{Flare analysis includes all photons after $\phi=0.85$ in Figure~\ref{fig:xray}.}
\label{tab:xray}
\end{deluxetable*}

\begin{deluxetable}{lc}
\tablewidth{0pt}
\tablecaption{Photometric Orbital Ephemeris of \mspthree}
\tablehead{
\colhead{Parameter} & \colhead{Value}
}
\startdata
R.A. (J2000)\tablenotemark{a}  & $08^{\rm h}38^{\rm m}50^{\rm s}\!.416$ \\
Decl. (J2000)\tablenotemark{a} & $-28^{\circ}27^{\prime}57^{\prime\prime}\!.03$ \\
Time span (MJD)           & 57071--58193 \\
Epoch $T_0$ (MJD TDB)\tablenotemark{b}      & 57781.2515(9) \\
Orbital period $P_{\rm orb}$ (day)  & 0.2145229(6) 
\enddata
\label{tab:ephem}
\tablenotetext{a}{Position from Pan-STARRS1 \citep{fle16}.}
\tablenotetext{b}{Epoch of ascending node of the putative pulsar,
$\phi=0$ in Figure~\ref{fig:optical_phase}.}
\end{deluxetable}

\subsection{X-ray Observations}

We obtained three observations of \msptwo\ with the \chandra\
Advanced Camera for Imaging and Spectroscopy (ACIS) S33 CCD,
each spanning slightly more than one pulsar orbit.   Details
are listed in Table~\ref{tab:xray} and the light curves are 
shown in Figure~\ref{fig:xray}.  The first two observations
have similar flux and a strongly modulated light curve with
a minimum centered at $\phi=0.25$, which is a common feature
among redbacks.  The third observation has a higher count
rate and is dominated by flaring.

The shape of the first two light curves is sometimes interpreted
as X-rays emitted by an intrabinary shock very close to the
companion being partly occulted at inferior conjunction of the
star \citep{bog11}.  However, such an effect would tend to
produce a double-peaked light curve, whereas the data are
statistically consistent with a broad, single peak.

Instead, beaming of synchrotron due to bulk relativistic
motion along the limbs of the intrabinary
shock may be the dominant effect on the light curve,
as modelled by \citet{rom16b} and \citet{wad17}.  In this case,
the nature of the light curve depends on the location and shape
of the intrabinary shock, which in turn is determined by the relative momentum
of the companion and pulsar winds.  There is some evidence that companion
wind may be the stronger one in \msptwo\ as its radio pulsations
are never detected during $0.02 < \phi <0.49$ \citep{den16},
almost half the orbit, and are sporadically eclipsed at other
phases as well.  Since the companion wind is responsible for
the radio absorption, this means that the intrabinary shock
may be wrapped around the pulsar rather than around the companion,
with the X-ray emission beamed toward the observer at $\phi=0.75$.
A broad, single X-ray peak could be seen at this phase if the
bulk velocities along the shock are mildly relativistic,
or the inclination angle of the orbit is not too large \citep{wad17,rom16b}.

Power-law spectral fits consistent with synchrotron emission
are given in Table~\ref{tab:xray},
where the $N_{\rm H}$ is either a free parameter or fixed
at the total Galactic value along the line of sight;
the choice makes little difference.
The count rates adjusted for effective exposure
at each orbital phase are not significantly different for
the first two (``quiescent'') observations.  While the flux appears
to have changed, this is mostly due to the additional time spent
at brighter orbital phases in the second observation.

In order to estimate the X-ray luminosity
we have to consider distance estimates based on the
dispersion measure (DM=16.65 pc cm$^{-2}$, \citealt{den16}).
Distance is uncertain for MSPs at high Galactic latitude,
like \mspone, because it depends
on the unknown scale height of the electron density distribution.
The distance to \mspone\ from the NE2001 electron density model of
\cite{cor02} is 0.7~kpc.  However, \cite{den16} calculated
that the companion star would significantly underfill its Roche lobe
at this distance, which is generally not the case for redbacks,
and argues for a larger value.  In fact, two revised models of the electron
density distribution that incorporate a thick disk \citep{sch12,yao17}
both predict a distance of $\approx2$~kpc.  Adopting this value,
the 0.3--8~keV X-ray flux for the two combined quiescent observations
in Table~\ref{tab:xray} corresponds to
$L_x = 2.8\times10^{31}\,(d/2\,{\rm kpc})^2$ erg~s$^{-1}$.

The third light curve is clearly dominated by flaring, with the
main flare beginning at $\phi=0.85$.  The inset in Figure~\ref{fig:xray},
which has 100~s bins, shows that the flare has structure on this
timescale.  The photon energies change over the flare; 
the first peak is harder than the second.  Even before
the main event there seems to be lower-level activity, with a small
peak at $\phi=0.5$ that is higher than the usual quiescent maximum at
$\phi=0.75$.  We made two spectral fits, one for the entire observation,
and another for the main flare from $\phi=0.85$ to the end of the
observation.  Both fits have harder spectra than the quiescent state,
as listed in Table~\ref{tab:xray}.  The flare has $\Gamma\approx1.2$ while
the quiescent observations have $\Gamma\approx1.5$, although these differ
at only the 1$\sigma$ level.

Note that the peak flare count rate of
$\approx0.14$~s$^{-1}$ is 33 times higher than the mean of
the quiescent observations.  This implies that the peak 0.3--8 keV
luminosity is $\approx9.4\times10^{32}\,(d/2\,{\rm kpc})^2$ erg~s$^{-1}$,
which is $\approx12$\% of the pulsar's spin-down power \citep{den16}.
Here we have corrected the apparent spin-down power for the
kinematic \citep{shk70} effect using the Gaia measured proper motion of
($\mu_{\alpha},\mu_{\delta})=(-16.28\pm1.00,-11.70\pm1.27)$ mas~yr$^{-1}$
\citep{gai16a,gai16b}, resulting in
$\dot E=[1.16-0.32(d/2\,{\rm kpc})]\times10^{34}$ erg~s$^{-1}$.

\begin{figure}
\centerline{
\includegraphics[angle=0,width=1.\linewidth,clip=]{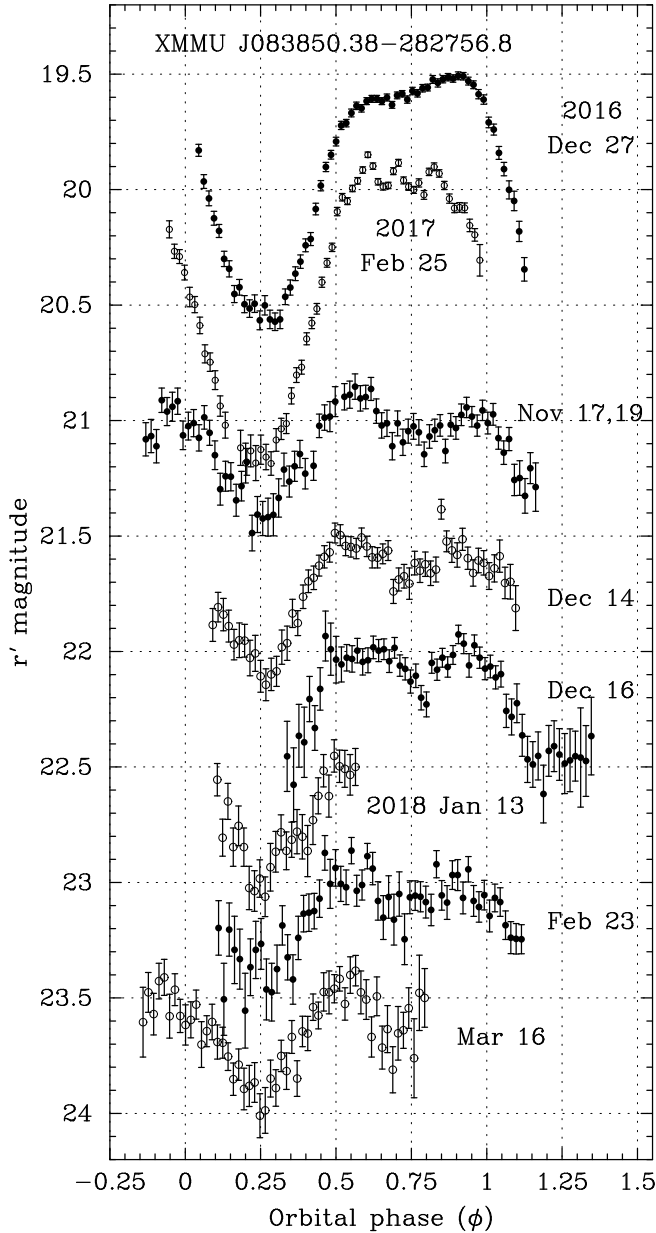}
}
\caption{Optical light curves of \mspthree\ 
as a function of orbital phase according to the ephemeris
in Table~\ref{tab:ephem}.   Previously published observations
from 2016 December and 2017 February are shown for comparison
with new data from 2017 November -- 2018 March.
A log of all the observations is
given in Table~\ref{tab:optlog}.  Each data set after 2016 December 27
is displaced downward by a multiple of 0.5 mag for clarity.
The different symbols aid in the separation of
closely spaced light curves.
}
\label{fig:optical_phase}
\end{figure}

\begin{figure}
\includegraphics[angle=270,width=1.\linewidth]{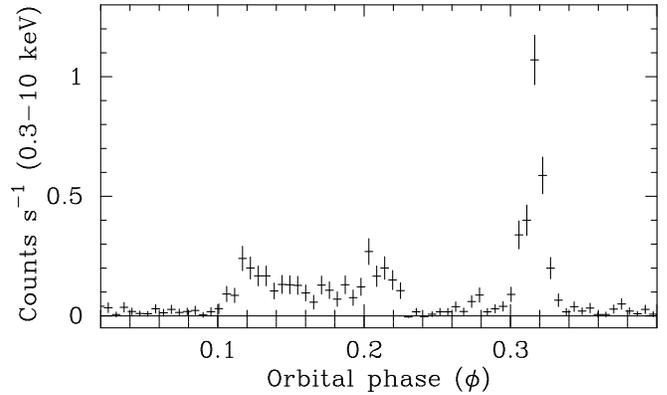}
\caption{Background-subtracted \xmm\ light curve of the 1.2~hr
 flaring episode of \mspthree\ on 2015 December~2 (ObsID 079080101),
 adapted from \citet{hal17b}.  The bin size is 100~s.  The revised
 orbital ephemeris of Table~\ref{tab:ephem} was used.}
\label{fig:xray_phase_fig.ps}
\end{figure}

\section{\mspthree}

\mspthree\ is a putative redback identified with \fgl.
Its 5.15~hr orbital light curve, as noted by \citet{hal17b},
bears a strong resemblance to that of \msptwo.  In
Figure \ref{fig:optical_phase} we show two of the six
light curves from the previous paper for comparison with
new data obtained in 2017 November -- 2018 March.
It is evident that in 2016 until at least 2017 March,
the star was in a flaring state very similar to that of \msptwo\
in Figure~\ref{fig:flare}, with rapid variability and brightness
on the heated side that is $\approx0.5$~mag higher than it was in
subsequent, quiescent periods.
There may be also be rapid variability during some of
quiescent episodes, particularly on 2017 December 14 and 16.

In the quiescent state of late 2017 -- 2018 a slight dip appears
around $\phi=0.75$, indicating that ellipsoidal
modulation may have a significant effect on the
shape of the light curve.
Similar to \msptwo, the minimum flux of \mspthree\
at $\phi=0.25$ has remained comparatively constant, at
$r^{\prime}\approx20.5\pm0.1$.  However, this is a fainter star,
sometimes difficult to measure accurately at minimum
because of its southerly declination (high airmass) and a bright,
neighboring star.  Its calibration is also less accurate,
with possible systematic variation of $\pm0.035$~mag.
Nevertheless the apparent upper limit of $\pm0.1$ mag
on the range of intrinsic stellar variability at $\phi=0.25$ 
is much less than the observed range of $\approx0.8$ mag
at the opposite phase of the orbit.
Thus, we conclude that variable heating and flaring is
confined to the side of the companion facing the pulsar.

With the new data, we were also able to fit a refined orbital
ephemeris using the method described in \citet{hal17b},
in which the epoch of minimum in the light curve
is used as the fiducial phase $\phi=0.25$.  The longer
baseline also enables us to extrapolate backward with a
precise cycle count to an observation obtained on
2015 February 18 (Figure~7 of \citealt{hal17a}).
This is the earliest known light curve of the star,
before its orbital period was apparent, but it contains
a dip which provides another timing for phase 0.25.
The resulting phase-connected ephemeris spans 3 years,
and improves the precision of the previously published orbital
period by an order of magnitude.
In Table~\ref{tab:ephem} we list the revised orbital parameters,
as well as the position from Pan-STARRS1 \citep{fle16}, which differs by
$0.\!^{\prime\prime}4$ from the position derived in \citet{hal17b}
based on USNO B1.0 astrometry.  

Using the new ephemeris, which spans the epoch of the
\xmm\ observation of 2015 December~2 (ObsID 079018010),
we are able to assign a precise orbital phase to the dramatic
1.2~hr long X-ray flare \citep{hal17b} seen in that light curve.
Figure~\ref{fig:xray_phase_fig.ps} shows
the flaring segment with its three peaks spanning $0.10<\phi<0.34$.
(An accompanying optical flare, observed with lower time resolution,
is not reproduced here).  Note that the phase has changed by 0.15
from the extrapolation of the former ephemeris that did not span
the epoch of the X-ray observation.
Interestingly, this event is almost centered
on inferior conjunction of the secondary star, unlike all of the
other flares reported here.  Because of the complex temporal structure,
it is not clear whether the quiescent level around $\phi=0.25$ is an
interval between spikes, or an eclipse by the secondary.

\section{\mspone}

\begin{figure*}
\includegraphics[angle=0,width=0.98\textwidth]{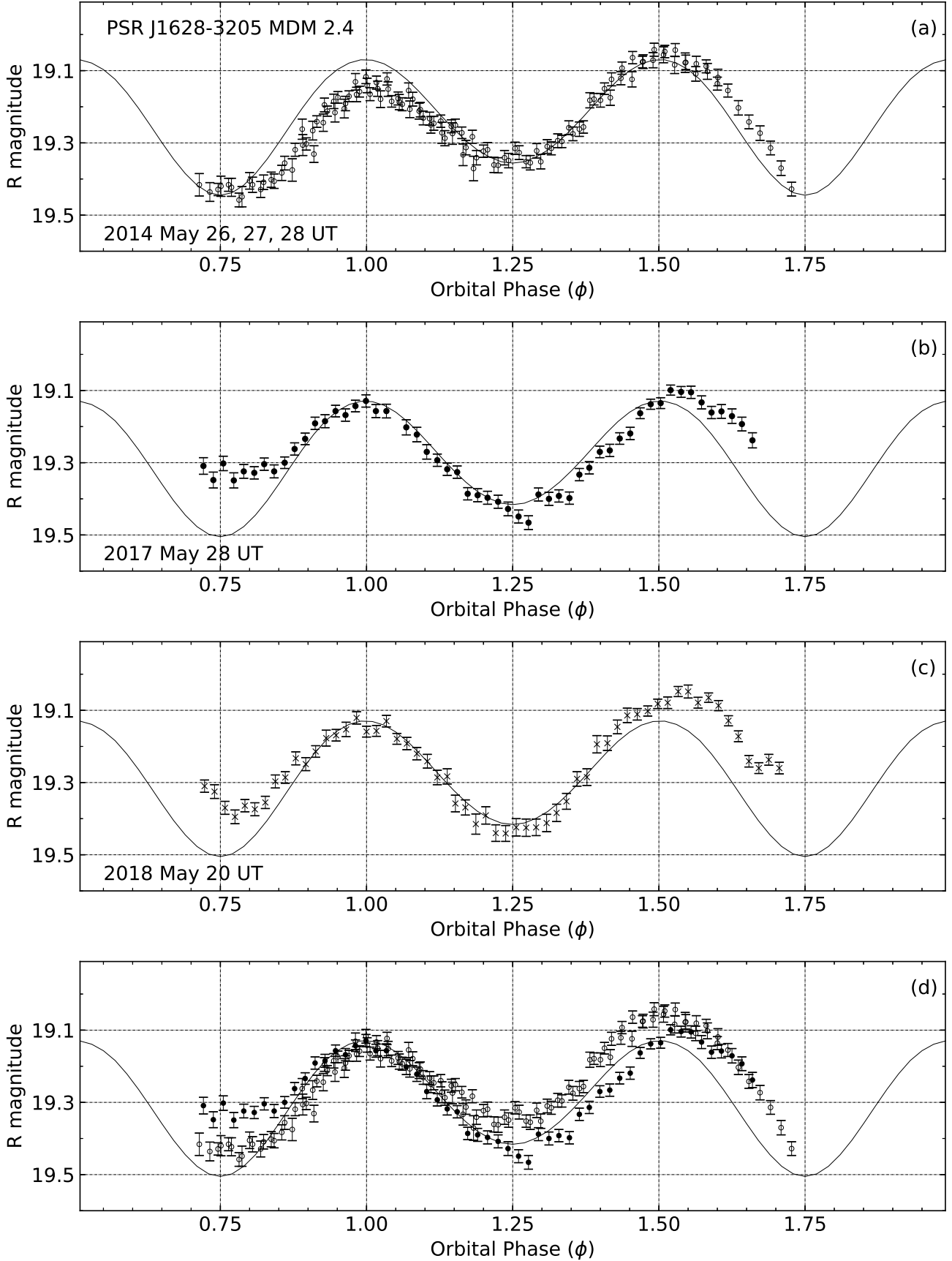}
\vspace{-0.1in}
\caption{$R$-band light curves of \mspone. 
 (a) three consecutive nights
 in 2014 May superposed, adapted from \citet{li14}.
 The curve is an ellipsoidal model from that paper,
 with inclination angle $i=75^{\circ}$.
 (b) a new $R$-band light curve obtained on 2017 May 28,
 with the same model curve displaced by 0.06~mag.
 (c) the latest $R$-band light curve,
 obtained on 2018 May 20, is very similar to the one from 2017.
 (d) comparison of 2014 and 2017 data. 
}
\label{fig:mspone_phase}
\end{figure*}

\mspone\ was first studied optically by \citet{li14} in 2014 May, where its
5.00~hr light curve appeared to be dominated by ellipsoidal modulation.
We fold the data on an unpublished orbital ephemeris that spans 3
years from 2010 November to 2013 November, with $T_0 = 55514.86519315(12)$ MJD,
$P_{\rm orb}=0.2081445829(9)$~d (S. Ransom 2014, private communication).
As before, we use only the leading term because
period derivatives, needed to fit slow wandering by tens of seconds
in the time of ascending node (e.g., \citealt{den16}), are small and
are not predictive outside the span of the ephemeris.

In Figure~\ref{fig:mspone_phase} we show the 2014 data with the best-fit
(by eye) model of ellipsoidal modulation from that study,
where only a lower limit on the inclination angle, $i>55^{\circ}$,
could be determined.  The representative model shown has $i=75^{\circ}$. 
Asymmetries in the data, deviating from the model, could only be explained
by invoking off-center heating effects, starspots, or magnetic
channeling of the pulsar wind.  In 2017 May we obtained an additional 
light curve for \mspone\ that covered almost a full orbital period
(also shown in Figure~\ref{fig:mspone_phase}).
We observe that the minimum at phase $\phi=0.25$ has become
$\approx~0.1$~mag fainter than in 2014, while the minimum at
$\phi=0.75$ has become $\approx~0.1$~mag brighter.
The maxima at phases $\phi=0.5$ and $\phi=0$ are still unequal,
and the minimum at $\phi=0.75$ and the maximum at
$\phi=0.5$ are delayed by about 0.05 in phase.  One more
observation was made in 2018 May (Figure~\ref{fig:mspone_phase}),
and the resulting light curve is little changed from 2017 May.
Systematic variation from year to year due to the calibration
process is limited to $\pm0.01$~mag, so the observed changes
must be real.

Thus, three changes can be discerned between 2014 and 2017.
First, the respective heights of the two minima in the light curve
are inverted.  In 2014, the interpretation of the minima was that 
pulsar heating is minimal.  But in 2017, the higher brightness at $\phi=0.75$
than at  $\phi=0.25$ suggests that pulsar heating has increased.
However, it is difficult to explain why the absolute level at $\phi=0.25$
decreased in 2017 unless in 2014 there was some pulsar wind being channeled
to the ``night'' side of the companion by its own magnetic field, an effect
that is no longer present in 2017.  In any case, it seems that variable
heating by the pulsar does play a role in this system.

Second, the unheated parts of the companion in 2017 seem to be on average
fainter than they were in 2014.  Although there is still some asymmetry in the
light curve, which was attributed in 2014 to a shock that is skewed toward the
trailing side of the companion, that effect is less pronounced in 2017.
The decrease in overall flux may suggest a decrease in radius of the companion.
The persistence of high-amplitude orbital modulation indicates that the
companion star is still tidally distorted, but it may have shrunk slightly.

The third change is that in 2017 the minimum at $\phi=0.75$,
and possibly the one at $\phi=0.25$, are delayed by about 0.05 in phase,
as is the maximum at $\phi=0.5$, while in 2014 only the minimum at
$\phi=0.75$ was lagging.
These changing phase shifts are further evidence of variability due to
asymmetric heating from the pulsar or other causes intrinsic to the companion
such as the distribution of large star-spots or magnetic activity cycles.

\section{Discussion}

All three systems studied here are redback MSPs in the non-accreting state,
as evidenced by their magnitudes, red colors, and substantial orbital
modulation.  They have low-mass, nearly Roche-lobe filling
companion stars that are tidally distorted and heated by the pulsar wind.
The light curves of redbacks can be classified according to whether
the modulation is dominated by pulsar wind heating or ellipsoidal geometry.
\msptwo\ and \mspthree\ are strongly heated, and exhibit significant changes
on time scales of weeks to months, which we interpret as structural variations
in the heating mechanism.  The little data that we have on \mspone\ also
shows variations in the heating pattern between observations obtained in
2014 and 2017, albeit at a lower level compared to its dominant
ellipsoidal modulation.

The light curves and variability of \msptwo\ and \mspthree\ are very
similar to each other.
Both have $\approx0.5-1$ mag orbital variation due to pulsar heating,
with the larger-amplitude episodes associated with rapid flaring.
Even when flaring is absent, the light curve is asymmetric about
the line between stars, usually sloping downward in the phase range
$0.5<\phi<1.0$. Very similar behavior \citep{wou04,tho05}
is seen from the transitional MSP J1023+0038 in its radio-pulsar state.
An asymmetric heating distribution produced by
an intrabinary shock distorted by orbital motion
could explain part of the modulation \citep{rom16b}. However, the
modeled asymmetries are not great enough to account for the observed
discrepancy between the brighter and fainter magnitudes of the slope.
Magnetic fields channeling pulsar wind particles and radiation directly
onto the surface of the companion have been suggested to account for
the remainder of the observed asymmetry \citep{li14,tan14,san17}. 

Repeated observations spaced by a few days to a few months allowed
us to investigate variable heating in detail. The largest changes
in the light curves occur at phase $\phi=0.75$ for \msptwo\ and \mspthree,
where pulsar heating is expected to be most visible.
On timescales of weeks or months, the brightness at this phase varies
by 0.2 to 0.4~mag, with an additional 0.5~mag increase when in a
flaring state.  Even in the faintest states of these two pulsars,
inferior conjunction of the companion ($\phi=0.25$) is always fainter
than superior conjunction ($\phi=0.75$),
indicating that pulsar heating is always important.  
From their preferred orbital phases, the optical flares appear to be located
on or near the heated face of the companion,
although a bright X-ray/optical flare was seen from \mspthree\ at the
opposite phase (Figure~\ref{fig:xray_phase_fig.ps}).  Also, there is
the case of PSR J1311$-$3430, a frequently flaring BW
that has flares at all phases \citep{an17}.

The X-ray flare seen in one of the three \chandra\ observations
of \msptwo\ has a peak luminosity of $\approx10^{33}$ erg~s$^{-1}$,
which is a significant fraction, $\approx12\%$ of the pulsar's spin-down power. 
The bolometric luminosity of the probable non-thermal spectrum is likely
to be even larger, which suggests that some stored energy in
magnetic field is being released.  The photon spectral index as
hard as $\Gamma=1.2$ in this flare, and in the persistent
emission of other redbacks, can
be taken as evidence that magnetic reconnection in a striped
pulsar wind, rather than shock acceleration, is responsible for
energizing synchrotron emitting electrons.
See the discussion in \citet{aln18} of this point, in the context
of the persistent X-ray emission of the redback PSR J2129$-$0429.
Since there is no simultaneous optical observation of this event,
it is difficult to know if it is associated with an optical flaring
episode, via direct emission or reprocessing.  However,
it seems that X-ray flares are common enough
in \msptwo\ and \mspthree\ that several could have occurred
during our $>30$ optical observations of these systems.

In 2.4 years of monitoring \msptwo, we have seen at least
two ``cycles'' in heating amplitude, although there may have been
more events in the interval from 2017 April to December when there were
no observations.  There is evidently some mechanism that modulates
the strength of the pulsar wind heating on a timescale of months.
The duration of the flaring state is difficult to pin down.  In
retrospect, it lasted at least 3 months (2016 December to 2017 March)
in \mspthree, which was the span of observations in \citet{hal17b},
but it had stopped when observations resumed 9 months later (2017 November).
For \msptwo\ we have only observed flaring during the three
(consecutive) nights of this program (2018 April 19--21), whereas there
was minimal heating and no flaring 2 months earlier and 1 month later.

\mspone\ displays the same sort of variable and asymmetric pulsar wind heating,
albeit at a lower level compared to the ellipsoidal modulation in this system.
The two maxima of the light curves are unequal, and the minimum at $\phi=0.75$ is
often higher than the one at $\phi=0.25$, both effects indicating a heating
contribution.  The phase delays of the $\phi=0.5$ (higher) maximum and the
$\phi=0.75$ minimum indicate that the heating has the same sense of asymmetry
that characterize \msptwo, \mspthree, and PSR J1023+0038.  In addition, there
is evidence for variable heating even at $\phi=0.25$ in \mspone.
This unexpected effect may be due to a process in which magnetic fields
intrinsic to the companion channel pulsar wind to localized areas \citep{san17},
including the ``night'' side of the companion.  Possible mechanisms include
a magnetic field which shifts in azimuth to channel the pulsar wind to 
different locations on the companion, or a large, migrating star spot.
Strong magnetic fields and large starspots can be expected on rapidly
rotating, tidally locked stars.  \citet{van16} found evidence for
differential rotation
of such spots during intensive monitoring of the light curve of the redback
PSR J1723$-$2737, which sometimes showed a period slightly different
from the orbital period.

Finally, we note that some of the more extreme light curve shapes and
variations that we observe may help to understand the optical data on other putative
redback MSPs for which a pulsar has not yet been detected and spectroscopic
radial velocities are not yet available to establish the orbital phase.
In particular, the double-peaked and asymmetric 5.47~hr orbital light curve of
the putative counterpart of 3FGL J2039.6$-$5618 \citep{rom15a,sal15} resembles
closely those of \mspone, and especially \msptwo\ during its epochs of weaker heating.
Indeed, \citet{sal15} had to develop a model involving asymmetric heating
along the lines discussed here to fit the optical light curve of 3FGL J2039.6$-$5618.
It will be interesting to see, when the absolute phasing of that system is accomplished,
if their phase assumptions were correct.

\section{Conclusions}

Comparisons among light curves collected over intervals of days to years 
show clear indications of variable heating in the three redback MSPs
observed.  The pulsar wind heating is generally off-center, manifested
as shifts in the intensity and phase of the extrema of the light curves
from expected symmetry based on the radio ephemerides.
Generally, the companion star is brighter at $\phi=0.5$ than at
$\phi=1.0$, which would imply that its trailing side is hotter
than its leading side.  The light curves
exhibit changes on time scales of weeks to months, which may be due
to changing magnetic fields intrinsic to the companion star channeling 
the pulsar wind to localized areas, or migrating star spots. 
The largest variations are centered around superior conjunction
of the secondary ($\phi=0.75$), and include episodes of decreased
heating as well as flaring. 

We suggest that these behaviors may be due to shifts in magnetic 
fields channeling the pulsar wind or non-synchronous rotation which
repeats on time scales of months to years.  Given the apparent cyclic
nature of the variability, it seems plausible that the mechanism
must be at least partly if not wholly due to the latter and may reflect
a combination of both. Continued optical observations will more
definitively characterize this episodic variability.
Further X-ray monitoring of these objects could also examine
the geometry and stability of the intrabinary shocks in more
detail, and investigate the locations of X-ray flares, whether in
the shock or on the companion.  Simultaneous X-ray and optical
spectroscopy of flares would be especially informative about their
origin.

\section{Acknowledgements}

This work is based on observations obtained at the MDM Observatory,
operated by Dartmouth College, Columbia University, The Ohio State University,
Ohio University, and the University of Michigan.  Support for this work was
provided by the National Aeronautics and Space Administration through
\chandra\ Award Number SAO G07-18047X issued by the
{\it Chandra X-ray Observatory} Center,
which is operated by the Smithsonian Astrophysical Observatory for and on
behalf of the National Aeronautics and Space Administration under contract
NAS8-03060.

\end{document}